# Students' Views of Macroscopic and Microscopic Energy in Physics and Biology

Benjamin W. Dreyfus, Edward F. Redish, and Jessica Watkins

*Department of Physics, University of Maryland, College Park, MD  20742*

**Abstract.** Energy concepts are fundamental across the sciences, yet these concepts can be fragmented along disciplinary boundaries, rather than integrated into a coherent whole.  To teach physics effectively to biology students, we need to understand students' disciplinary perspectives.  We present interview data from an undergraduate student who displays multiple stances towards the concept of energy.  At times he views energy in macroscopic contexts as a separate entity from energy in microscopic (particularly biological) contexts, while at other times he uses macroscopic physics phenomena as productive analogies for understanding energy in the microscopic biological context, and he reasons about energy transformations between the microscopic and macroscopic scales.  This case study displays preliminary evidence for the context dependence of students' ability to translate energy concepts across scientific disciplines.  This points to challenges that must be taken into account in developing curricula for biology students that integrate physics and biology concepts.

## INTRODUCTION

Recent reports on the reform of undergraduate education for future physicians [1] and biologists [2] emphasize the need for deeper integration of physical principles into biology education.  Energy is a particular focus of this integration, as a physical concept that is fundamental to biology.  Ref. 2 includes energy concepts in a list of "central themes" of biology, and Ref. 1 includes energy concepts in its learning objectives connected to physics, chemistry, and various scales of biology.

Energy is unusual among basic physics concepts in that even experts disagree about its fundamental nature.  Thus, there has been vigorous debate in the physics education and science education literature not only about the learning and teaching of energy [3-7], but about how energy should be understood. [8-13] For example, there is discussion about whether energy should be ascribed substance-like properties or whether it is an abstract concept defined only in terms of conservation; whether "the ability to do work" is a useful perspective for understanding energy; and whether the transformation among "forms of energy" is a coherent concept.  Very little of the previous research on student understanding or teaching has focused specifically on energy in biology or on the disciplinary interface between physics and biology, but some have raised red flags around this.  Trumper [14] shows that pre-service biology teachers have confusion around physics and specifically energy.  Gayford [15] writes that "for those who do combine their study of natural science with physical science, the ideas that they are taught about energy appear remote from what occurs in biological systems," and Lin [16] goes further to show compartmentalization (with regard to energy) among different hierarchical levels within biology.

At the University of Maryland we are piloting a new physics sequence for biology majors in 2011-12 with a goal of achieving stronger integration of physics with biology.  A major goal in developing this course is to investigate how energy concepts can be taught in a way that unifies the way energy is used in physics, biology, and chemistry, and that transcends the disciplinary barriers.  We expect that disciplinary differences in the use of the energy concept may confuse students the way it is currently taught, and the case study presented here supports this hypothesis.  In this paper we look at how one student does and does not connect physics and biology, and the microscopic and macroscopic scales, in understanding energy.

## METHODS

In order to get a sense of how our biology majors thought about the concept of energy, we interviewed five undergraduate students, all of whom were enrolled in the first semester of algebra-based introductory physics, and all of whom were life science majors and/or pre-health-care students. In this paper we look at one student, "Dennis," a junior who had recently switched his major from biology to ecological technology, but had taken all of the courses in the introductory biology and chemistry sequences and was completing the pre-med requirements. We conducted two interviews with Dennis, one in the middle of the semester (before his physics class had studied energy) and one at the end of the semester.

The interview protocols were designed to elicit student understanding of certain conceptual issues connected to energy in biology and chemistry which are not addressed in this paper. But in Dennis's interviews, an interesting pattern emerged in the data, initially unprompted. This pattern highlighted epistemological attitudes about energy, particularly on the divisions between physics and biology and between the macroscopic and microscopic scales. Because this was not part of the original interview protocol, this line of discussion does not appear substantially in the other students' interviews.

Because this paper is based on one student, we are not making claims about the entire population of students, but we believe these data are still instructive in bringing up issues that will inform future research and curriculum development.

## PERCEIVED DISCONNECT BETWEEN PHYSICS AND BIOLOGY

At the conclusion of two interviews about energy, Dennis was asked whether he perceived any differences between how energy was approached in his physics class and how it was approached in biology and chemistry. He responded that his physics class

> "talks a lot more about physical objects, stuff like that, which you don't really talk about in bio or chem. You don't really talk about macro stuff, you kind of talk about like interactions of molecules, in biology you talk about-- Chemistry, you talk more about interactions of like atoms and stuff like that. Biology, it's more about interactions of molecules."

He later confirmed that by "physical objects" he meant macroscopic objects. Thus he discusses energy in physics as dealing with the macroscopic scale and energy in biology as dealing primarily with the microscopic scale. This makes sense, since in the traditional first-semester physics class in which he was enrolled, only macroscopic phenomena were considered.

When discussing various microscopic phenomena such as photosynthesis, Dennis describes energy in terms of electrons (as discussed below). When asked if this has any relationship to kinetic and potential energy (which Dennis had brought up earlier in the interview), he says in various ways that these two types of energy (microscopic and macroscopic) are not directly comparable. One approach he takes is to say that they have "different units":

> "For instance, kinetic energy is measured in terms of like mass and volume [sic], and potential energy is mass, gravity, and height ... Whereas energy like in, you know, a chemical equation or something like that is, ... when we're doing redox reactions, it's energy potential of the chemical equation is measured in volts."

At other times, Dennis allows for the possibility that it may be possible to make connections between the different scales of energy, but says he does not find it useful:

> "Like assuming electrons are measured in, or you know, electrical charge or something like that. It's measured in like volts. ... And then when you're measuring movement and stuff like that of actual, like, of larger bodies, you use units like force, and stuff like that. And maybe you could convert the two, between the two, but I don't really see the point. ... I'm saying even if there were a way to connect the two, which I don't, I certainly don't, can't think of a way, I don't really think there would be a point in doing so."

In some cases, he seems to be making the claim, as expert scientists would do, that when we study phenomena at a particular scale, it is most practical to ignore irrelevant phenomena at other scales:

> "I'm sure you could describe this [picking up an object on the table and dropping it] at a chemical level or something, a molecular level for the aggregate of, you know, all the molecules, but I don't think that would be particularly helpful or useful.... But we generally don't use velocity or height, you know, to determine, to discuss molecular interactions, we just talk about a different set of units that we find more helpful and descriptive."

But in other instances, he suggests that a more fundamental distinction is in play. He refers to *"a situational use of the term energy,"* implying that energy is not a unified entity that exists at different scales, but only a **term** that can be used, by analogy, in different situations (in the same way that one might say "I don't have the energy to do this now" without intending to invoke any technical definition of energy).

Does Dennis see the distinction between macroscopic and microscopic energy as pragmatic or as more fundamental? He may not have one coherent answer to that question. But either way, he expresses the idea that these are distinct. And because he associates macroscopic energy with physics and microscopic energy with biology, this is tied to a disconnect between physics and biology in regard to energy.

## SUCCESSFUL BRIDGING BETWEEN SCALES AND DISCIPLINES

Yet despite these barriers between microscopic and macroscopic energy, and between physics and biology, Dennis successfully reasons across these barriers under a number of circumstances. Here we examine the circumstances that make this possible.

One type of instance in which Dennis connects the macroscopic/"physics" context to the microscopic/biology context is when this connection is merely an **analogy**. For example, when explaining how energy is stored in ATP, he says:

> "So this is ADP and this is P, the bond between these two, these phosphorus, it's really strong in that this is really strong negative charges, so you push those suckers together, it's hard to do that, but if you do that, then you have a whole lot of potential energy, because you know, when two molecules are, you know, kind of like magnets. If you shove two magnets together, you know, they have a whole lot of potential energy just 'cause, or pushing in a spring even, same deal, you know, you have a whole lot of potential energy, and as soon as you release that potential energy, the spring expands again. That's how work is done."[1]

Dennis sees the repulsive force between magnets and the elastic potential energy in a spring as productive analogies for understanding the mechanism for energy changes in a chemical reaction. It is unlikely that he thinks that there are actual springs in the ATP molecule, but he finds this to be a useful metaphor in the same way that a biologist might explain evolution in terms of selective pressure (without implying that this "pressure" is a force per unit area).

Here, we can distinguish between a recognition that two phenomena at different scales share an analogous structure (as in this case) and a recognition that the two phenomena are physically related.

Dennis also connects biological phenomena to "physics" (i.e. non-biological) phenomena when they are both at the microscopic scale. When asked about the different forms that energy can take, he arrives at the conclusion, seemingly on the spot, that electrons are the energy carrier that unifies disparate microscopic energy phenomena:

> "I guess it would be electrons, is where energy is stored, I guess would be the moral of the story. Yeah. 'Cause I mean if you look at redox reactions, that's, you know, the movement of electrons. Photosynthesis, you know, you plug in a photon and, you know, you essentially plug in an electron, it bumps up a state. And you know, solar power, it's the same thing, the sun's photons hit the solar power, you know, it bumps it up, it catches the current, it goes through a circuit. That's what creates the energy. So I guess electrons would kind of be the current. The currency."

In the same interview in which Dennis says he cannot think of a way to convert between microscopic and macroscopic energy, he explains some phenomena in terms of microscopic energy converted to macroscopic energy. It is notable that these examples come up in a non-biological, "physics" context:

> "In a car, it's kind of a mini-explosion every time the spark plug ignites ... The bonds, I guess you're breaking the bonds of—that are stored in the gasoline. And the breaking of that bond, you know, turned the energy, you know, I guess the, it's the energy released by that, but anyways, the breaking of that bond is what turns the piston."

He uses this to answer a question about how a cannon works:

> "So I guess in the same way, with a cannon, you ignite it, and you break the bonds that, I guess, have a whole lot of energy stored up, 'cause that's what makes them explosive material as you break them, it converts the energy of that to a cannonball. Or to pushing the cannonball, and then the cannonball moves. So I guess energy is kind of imparted from explosive material to the cannonball. From, I don't know if it'd be thermal, but whatever energy, whatever you'd call that energy stored up, I guess chemical energy, it gets converted into kinetic energy and that's a cannonball moving."

Here he talks about energy in chemical reactions and (macroscopic) kinetic energy, the very examples he cited as having "different units" and therefore being effectively incommensurable.

---
[1] We present this data without comment on whether Dennis's model is a correct description. The issue of energy in chemical bonds is complex in regard to both student understanding and instruction [17-18], and will be the subject of a future paper.

Absent from the data, however, is an example of connecting microscopic biological phenomena to macroscopic non-biological phenomena. While this absence does not prove anything, it suggests that for Dennis, bridging **either** the physics/biology or the microscopic/macroscopic barrier is easier than bridging both at the same time.

## DISCUSSION

These data show a student whose understanding of energy includes a disconnect in the abstract between the microscopic and macroscopic scales and between the disciplines of physics and biology, but who can make connections across these barriers when it is useful in understanding specific situations. Dennis may not have an explicit epistemological commitment to energy as a unifying principle in the same way that an expert would. He integrates energy across different scales and disciplines when the context facilitates this, and otherwise expresses little use for this integration. As recent work [19] has shown, students' epistemological stances can shift in response to context, and Dennis's shifting stance on energy provides an instance of this.

If it is an educational goal for students to integrate energy concepts across these domains, then these data show us that even a student with a reasonably sophisticated conceptual understanding of energy may not do that integration spontaneously under ordinary instruction, but that explicit efforts are necessary.

For physics classes, this can take the form of a more detailed treatment of chemical energy, including an understanding of the mechanisms by which energy is converted among different forms. The standard physics curriculum includes such a mechanism for conversion of (macroscopic) mechanical energy between kinetic and potential energy (work is done by a force), but "chemical energy" is treated as a nebulous "other" category. Making it clear that "chemical energy" (which the students know from their biology and chemistry classes) **is** kinetic and potential energy, at a smaller scale, can help students understand that energy is the same entity that exists continuously at all scales.

Biology classes spend considerable time on how photosynthesis and respiration result in the synthesis of ATP, but less time on how the energy stored via ATP is used. They could include a more explicit focus on how these microscopic mechanisms relate to macroscopic processes.

Evaluating the new physics curriculum for biology students should include assessments of how well students can integrate energy concepts between physics and biology and between different scales, in comparison to students with a traditional, less integrated curriculum.

This is only the beginning of our investigations into energy and thermodynamics in physics and biology. Despite all the philosophical and pedagogical complications, energy conservation and transformation in both physics and biology are relatively well understood. Directions for future research will have to include thermodynamic concepts such as entropy [20] and free energy [21], on which less work has been done.

## ACKNOWLEDGEMENTS

The authors wish to thank the University of Maryland Physics Education Research Group and Biology Education Research Group. This work was supported in part by NSF grants 09-19816 and DGE0750616.

## REFERENCES

1. Scientific Foundations for Future Physicians: Report of the AAMC-HHMI Committee (AAMC/HHMI, 2009).
2. National Research Council (US). Committee on Undergraduate Biology Education to Prepare Research Scientists for the 21st Century, Bio 2010: Transforming Undergraduate Education for Future Research Biologists (Natl Academy Pr, 2003).
3. D.M. Watts, Physics Education **18**, 213 (1983).
4. J. Solomon, Physics Education **20**, 165 (1985).
5. J. Solomon, Int. J. of Sc. Educ. **5**, 49-59 (1983).
6. N. Papadouris, C.P. Constantinou, and T. Kyratsi, J. Res. Sci. Teach. **45**, 444-469 (2008).
7. J. Solbes, J. Guisasola, and F. Tarín, J Sci Educ Technol **18**, 265-274 (2009).
8. R.L. Lehrman, The Physics Teacher **11**, 15-18 (1973).
9. N. Hicks, The Physics Teacher **21**, 529-530 (1983).
10. R. Duit, Int. J. of Sc. Educ. **3**, 291-301 (1981).
11. J.W. Warren, Int. J. of Sc. Educ. **4**, 295-297 (1982).
12. R. Duit, Int. J. of Sc. Educ. **9**, 139-145 (1987).
13. W. Kaper and M. Goedhart, Int. J. of Sc. Educ. **24**, 81-95 (2002).
14. R. Trumper, Int. J. of Sc. Educ. **19**, 31-46 (1997).
15. C.G. Gayford, Int. J. of Sc. Educ. **8**, 443-450 (1986).
16. C.-Y. Lin and R. Hu, Int. J. of Sc. Educ. **25**, 1529-1544 (2003).
17. S. Novick, J. of Biol. Educ. **10**, 116-118 (1976).
18. H.K. Boo, J. Res. Sci. Teach. **35**, 569-581 (1998).
19. A. Gupta and A. Elby, Int. J. of Sc. Educ. 1-26 (2011).
20. R.H. Swendsen, Am. J. Phys. **79**, 342 (2011).
21. E.M. Carson and J.R. Watson, University Chemistry Education **6**, 4-12 (2002).